\documentclass[12pt]{article}

\usepackage{arxiv}

\usepackage[utf8]{inputenc} 
\usepackage[T1]{fontenc}    
\usepackage{hyperref}       
\usepackage{url}            
\usepackage{booktabs}       
\usepackage{amsmath,amssymb,amsfonts}
\usepackage{nicefrac}       
\usepackage{microtype}      
\usepackage{lipsum}		
\usepackage{graphicx}
\usepackage[square,numbers]{natbib}
\usepackage{doi}
\usepackage{algorithmic}
\usepackage{textcomp}
\usepackage{bm}
\usepackage{xcolor}

\definecolor{darkgreen}{RGB}{0, 128, 48}

\usepackage[labelfont=bf]{caption}
\title{Low Complexity Point Tracking of the Myocardium in
2D Echocardiography}

\date{} 					

\author{Artem Chernyshov$^{1,2}$, John Nyberg$^{1,2}$, Vegard Holmstrøm$^{1,3}$, Md Abulkalam Azad$^{1,3}$,\\[0.02in] Bjørnar Grenne$^{1,3}$, Håvard Dalen$^{1,3}$, Svein Arne Aase$^4$, Lasse Lovstakken$^{1,5}$, Andreas Østvik$^{1,3,6}$}
\affiliations{$^1$ Norwegian University of Science and Technology, Trondheim, Norway\\
$^2$ ProCardio Center for Innovation, Oslo University Hospital, Oslo, Norway\\
$^3$ Clinic of Cardiology, St. Olavs Hospital, Trondheim, Norway\\
$^4$ GE Vingmed Ultrasound AS, Horten, Norway\\
$^5$ Centre for Innovative Ultrasound Solutions, Trondheim, Norway\\
$^6$ SINTEF Digital, Department of Health Research, Trondheim, Norway}

\renewcommand{\headeright}{}
\renewcommand{\undertitle}{}
\renewcommand{\shorttitle}{}

\begin{document}
\maketitle

\begin{abstract}
Deep learning methods for point tracking are applicable in 2D echocardiography, but do not yet take advantage of domain specifics that enable extremely fast and efficient configurations. We developed MyoTracker, a low-complexity architecture (0.3M parameters) for point tracking in echocardiography. It builds on the CoTracker2 architecture by simplifying its components and extending the temporal context to provide point predictions for the entire sequence in a single step. We applied MyoTracker to the right ventricular (RV) myocardium in RV-focused recordings and compared the results with those of CoTracker2 and EchoTracker, another specialized point tracking architecture for echocardiography. MyoTracker achieved the lowest average point trajectory error at 2.00$\pm$0.53 mm. Calculating RV Free Wall Strain (RV FWS) using MyoTracker’s point predictions resulted in a -0.3\% bias with 95\% limits of agreement from -6.1\% to 5.4\% compared to reference values from commercial software. This range falls within the interobserver variability reported in previous studies. The limits of agreement were wider for both CoTracker2 and EchoTracker, worse than the interobserver variability. At inference, MyoTracker used 67\% less GPU memory than CoTracker2 and 84\% less than EchoTracker on large sequences (100 frames). MyoTracker was 74 times faster during inference than CoTracker2 and 11 times faster than EchoTracker with our setup. Maintaining the entire sequence in the temporal context was the greatest contributor to MyoTracker's accuracy. Slight additional gains can be made by re-enabling iterative refinement, at the cost of longer processing time. MyoTracker source code: \href{https://github.com/artemcher/myotracker}{https://github.com/artemcher/myotracker}
\end{abstract}

\section{Introduction}
Speckle-tracking echocardiography (STE) is an affordable and non-invasive method for assessing myocardial deformation. It is a more accessible and angle-independent alternative to Tissue Doppler Imaging \cite{koopman_comparison_2010}. STE works purely as a computer vision algorithm for B-mode recordings, where it tracks the interference patterns of ultrasound waves with the myocardium \cite{leitman_two-dimensional_2004}. These patterns, "speckles", are assumed to be mostly stable under small displacements over time. Currently, commercial solutions for echocardiogram analysis rely on block matching algorithms for STE, with possible vendor-specific adjustments. Other types of algorithms have also been proposed by researchers, such as elastic image registration and various optical flow estimation techniques \cite{heyde_elastic_2013, alessandrini_myocardial_2013}. Whichever tracking algorithm is used, the main objective is to estimate myocardial strain. Often, it is either the global or regional longitudinal strain (GLS or RLS).

\begin{figure*}
    \centering
    \includegraphics[width=\textwidth]{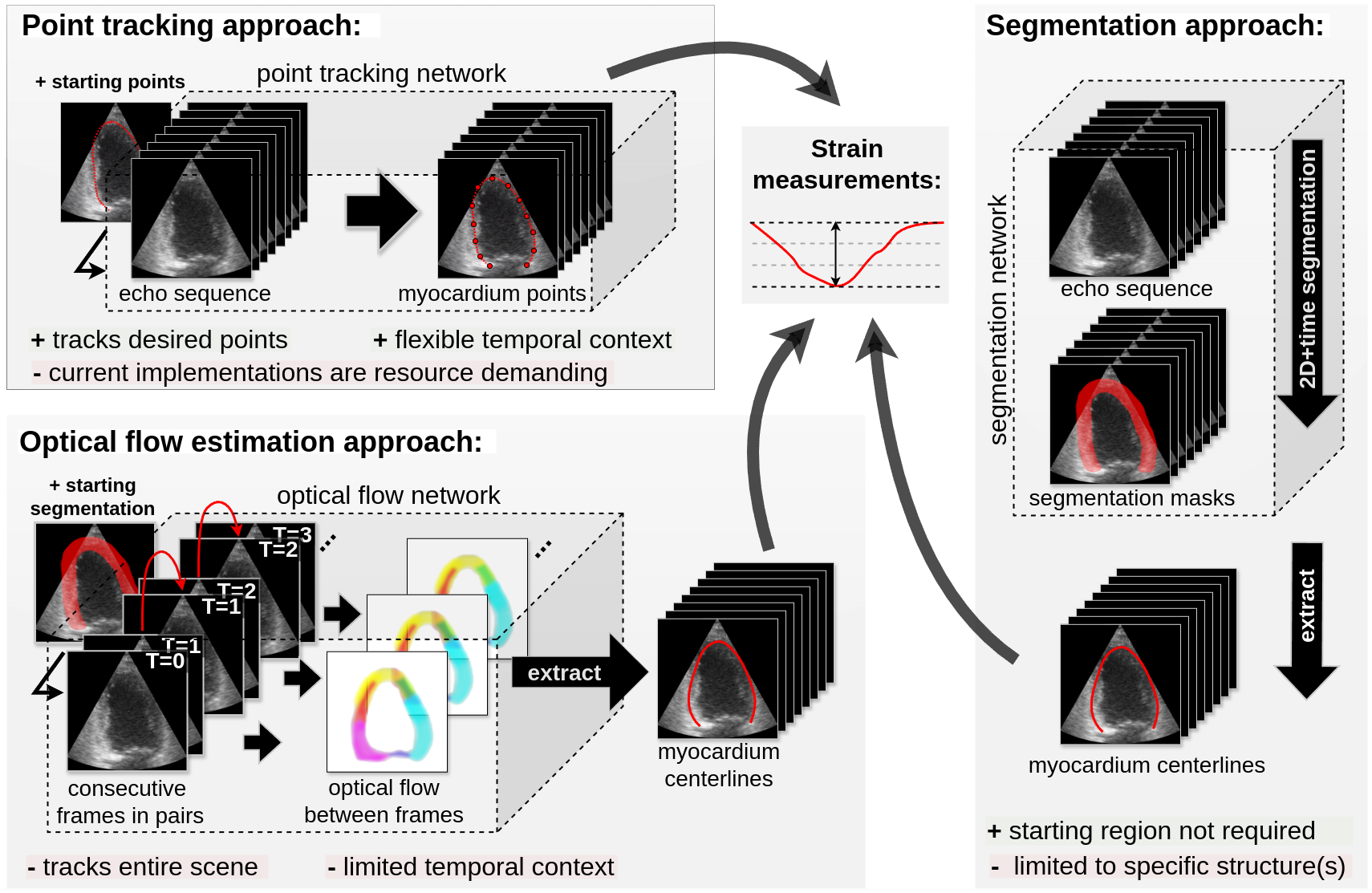}
    \caption{Deep learning approaches as alternatives to traditional STE. Point tracking and optical flow estimation are pure tracking methodologies for learning motion patterns, whereas segmentation methods detect specific structure(s) in every frame. Obtaining the myocardial outline in every frame of the video with any of the approaches allows strain calculations. The flexibility of point tracking methods allows for tracking only the desired points over variable-length sequences.}
    \label{fig:tracking_overview}
\end{figure*}

Both block matching and other traditional computer vision algorithms have significant drawbacks. In general settings, the lack of learning capabilities makes them rigid and sometimes unable to handle noise or occlusion. In echocardiography specifically, the challenging factors can include poor image quality, out-of-plane motion, foreshortening, and high heart rate. The resulting speckle decorrelation is something traditional STE algorithms struggle to overcome with their low adaptability and short temporal context windows (two frames) \cite{sitia_speckle_2010}. Although the details of their algorithms are unknown, the vendors that offer STE appear to combat these issues by imposing shape priors and varying degrees of regularization \cite{dhooge_two-dimensional_2016}. Such algorithm adjustments are likely manual and difficult to implement in a way that accommodates the sheer variety in the real cases. STE is only available in analysis software and not on the ultrasound systems in real time, which may suggest high computational demands as well.

\subsection{Deep learning algorithms for tracking}
The shortcomings of traditional algorithms can be addressed by deep learning algorithms, as they are capable of learning from data and detecting hidden patterns. The existing deep learning architectures for tracking can be broadly divided into dense optical flow estimation and point tracking categories, with the latter being more novel. Optical flow estimation architectures, such as FlowNet and the more advanced PWC-Net, track every pixel in a video from frame to frame \cite{dosovitskiy_flownet_2015,sun_pwc-net_2018}. Novel proposals using transformers or diffusion models have improved the accuracy of earlier studies \cite{huang_flowformer_2022, luo_flowdiffuser_2024}. However, architectures for optical flow typically have a limited temporal context, and new proposals are only beginning to overcome this constraint. In comparison, point tracking architectures can track only selected points. This approach is more efficient and allows for processing of more than two frames at once. TAP-Net \cite{doersch_tap-vid_2023} and PIPs \cite{harley_particle_2022} are considered initial attempts in this area, with PIPs++ \cite{zheng_pointodyssey_2023} being an improvement over the former and TAPIR \cite{doersch_tapir_2023} incorporating elements of both. CoTracker \cite{karaev_cotracker_2023} is one of the most recent architectures in the series. It outperforms its predecessors through "joint tracking," wherein it utilizes the correlations between point trajectories. Joint tracking would be a particularly beneficial property for an STE algorithm, as myocardium regions are jointly moving parts of a singular structure and not independent objects.

Deep learning algorithms have already been considered for STE, including optical flow estimation, point tracking, and temporally consistent segmentation approaches (Figure \ref{fig:tracking_overview}). The proposed algorithms were accurate and potentially more robust than traditional algorithms, but their computational efficiency remained suboptimal. For example, a modified PWC-Net was used in studies to track the left ventricular (LV) myocardium better than open non-learning optical flow algorithms \cite{ostvik_myocardial_2021, evain_motion_2022}. However, as an optical flow estimation method, it only processes two frames at a time and tracks the entire scene rather than just the necessary regions. Strain calculations then require at least one more non-trivial step of extracting the myocardial centerline from the displacement field. Point tracking has also been tried with a custom architecture made for echocardiography, EchoTracker \cite{azad_echotracker_2024}. The approach was more efficient than optical flow estimation, as it supported the option of tracking only the points placed on the myocardium. EchoTracker aimed to improve on CoTracker and preceding solutions by processing the entire recording at once, instead of in chunks of 5–8 frames or more. This decision increased the processing speed, but the architecture’s components remained as computationally intensive as those in general-purpose tracking architectures. As an alternative to pure feature tracking, segmentation methods have been used to detect the myocardium in a temporally consistent manner within each frame \cite{ouyang_video-based_2020, wei_temporal-consistent_2020, painchaud_echocardiography_2022, deng_myocardial_2022}. These methods require only the echo sequence as input, but their advantage is offset by reliance on complex convolutional architectures (over 10M parameters) or additional post-processing.

Reducing the complexity of deep learning algorithms to handle only echocardiography data may offer greater efficiency with little or no reduction in accuracy. This concept has already been validated on various heart chamber segmentation tasks with architectures limited to 0.3–0.4M parameters \cite{hu_automated_2024, chernyshov_automated_2024}. Compared to general visual data, echocardiography recordings are less varied. They only display cardiac structures that continually alternate between contraction and relaxation. The probe, serving as the camera, remains near-stationary. Regions of the myocardium move jointly in a relatively predictable pattern and with limited speed between frames. Thus, the complexity of the task is much lower compared to that of tracking in general visual scenes, even though speckle decorrelation, poor image quality, and morphological variability present certain challenges. Thus, the complexity and knowledge capacity of a deep learning STE algorithm can be significantly lower.

\subsection{Tracking the Right Ventricle}
The right ventricle (RV) has been less studied than the left ventricle (LV), but has gained increasing attention in recent years \cite{rudski_guidelines_2010, kossaify_echocardiographic_2015}. The chamber is more difficult to image with 2D echocardiography due to its anterior position in the chest, greater anatomical variability, complex geometry, and thinner walls. These properties lead to greater variability between recordings and poorer image quality. Nonetheless, 2D echocardiography remains the primary modality for RV imaging.

Difficulties in imaging can lead to complications in analyzing and tracking the RV. Validation of commercial tools for RV STE suggests that deformation imaging can still be performed on the RV myocardium, but the feasibility of measuring RV Free Wall Strain (RV FWS) is lower (83\%) than that of LV GLS (96\%) \cite{mcerlane_feasibility_2023, nyberg_echocardiographic_2023}. Beyond commercial initiatives, there have been very few studies on automating RV analysis and measurements in 2D echocardiography, and none that focus specifically on RV tracking.

\subsection{Aims and Contributions}
We aim to develop a deep learning algorithm for tracking the myocardium in 2D echocardiography that is accurate, fast, and less resource-demanding than state-of-the-art alternatives. The algorithm is designed and tested primarily for tracking the RV myocardium and consequent RV FWS measurements. Compared to LV tracking, RV tracking remains a more challenging task that fills a niche that is not as well covered by traditional STE or deep learning algorithms. Our proposed method, MyoTracker, uses the CoTracker point tracking architecture series (specifically, CoTracker2) as the baseline for its desirable properties, such as joint tracking. We develop MyoTracker from CoTracker2 through an extensive optimization process that expands its temporal context while simplifying or removing components deemed redundant. Thus, our contributions include:
\begin{itemize}
\item A novel application of deep learning in 2D echocardiography, as RV tracking has not yet been attempted or evaluated with such methods to date. Both RV tracking and RV FWS estimation are considered more difficult tasks than LV tracking and LV GLS measurements, respectively.
\item The MyoTracker architecture, which achieves greater RV tracking accuracy and RV FWS estimation accuracy than deep learning architectures previously identified as state-of-the-art for LV tracking.
\item A demonstration of the viability of low-complexity architectures for 2D echocardiogram analysis and quantification. The MyoTracker architecture features only 0.32M parameters, while requiring only 1 GB of GPU memory and 20–30 ms to process a whole cardiac cycle. These characteristics are several times or orders of magnitude better than those of the available alternatives.
\end{itemize}

\begin{figure*}
    \centering
    \includegraphics[width=0.95\textwidth, height=0.9\textheight]{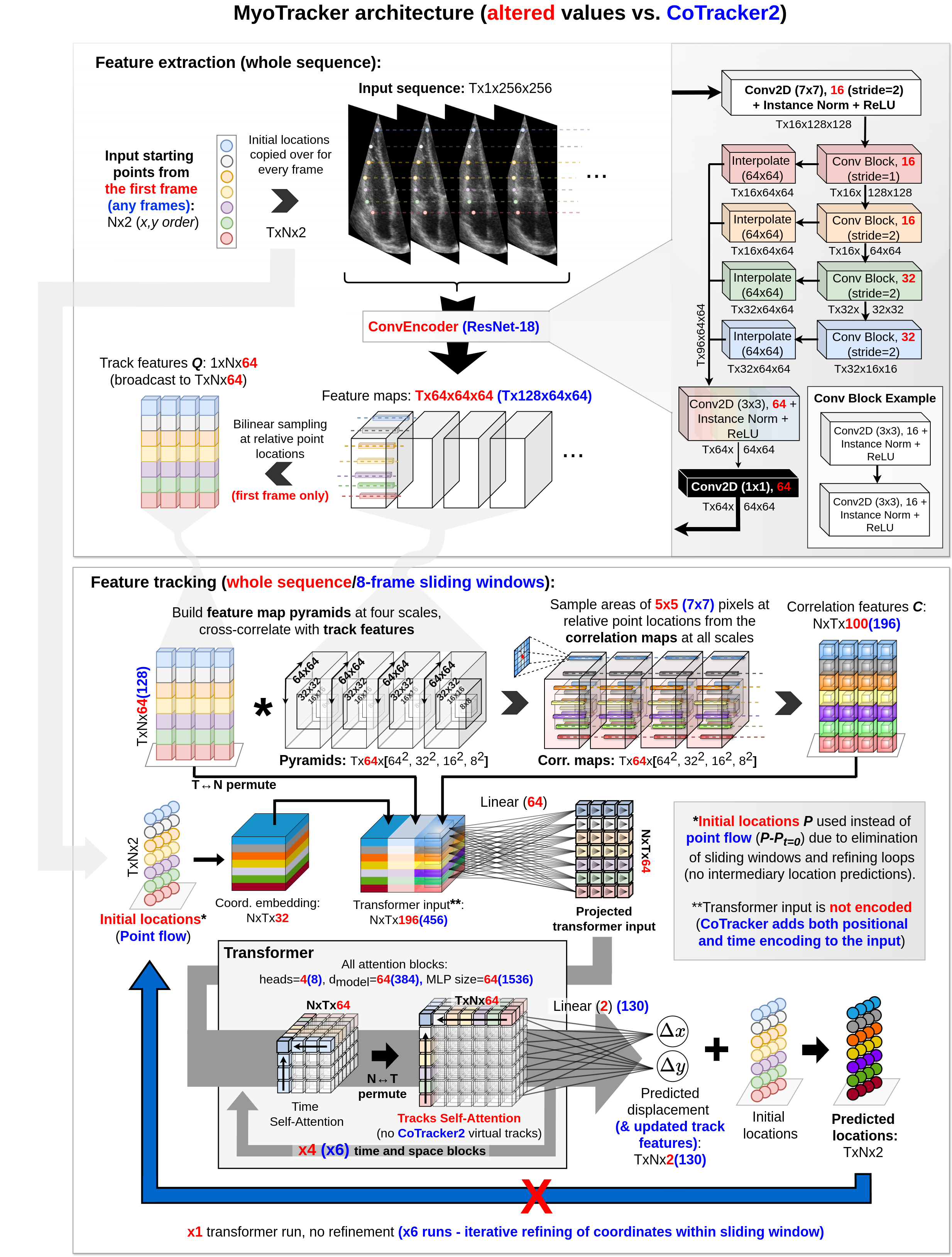}
    \caption{MyoTracker architecture. Highlighted in red are the elements that were altered from the definition of CoTracker2 \cite{karaev_cotracker_2023} to produce MyoTracker. Where possible, the original values are highlighted in blue. The point locations are predicted directly for all frames in the sequence, without iterative refinement. All CoTracker2 components pertaining to visibility estimation, sliding window processing, transformer input encoding, and virtual tracks were removed. These components are not included in the figure.}
    \label{fig:myotracker}
\end{figure*}

\section{Methods}
Our tracking method, MyoTracker, is a simplified version of the CoTracker2 architecture that inherits design choices from EchoTracker. We applied MyoTracker, baseline CoTracker2, and EchoTracker to RV-focused echo recordings. The training and evaluation data were obtained by processing the recordings with commercial clinical software and extracting the cardiac cycles, tracking points, and reference RV Free Wall Strain (RV FWS) values. The outputs (tracked points) from the models were provided to a separate algorithm to calculate RV FWS.

\subsection{CoTracker and EchoTracker Architectures}
CoTracker is a state-of-the-art, transformer-based architecture developed by Meta AI \cite{karaev_cotracker_2023}. It can jointly track a large number of points $\mathbf{P}$ across videos of arbitrary length $T$, estimating both the point coordinates and visibility. CoTracker processes all frames through a convolutional encoder (a ResNet-18 variant) to extract frame features, which are then sampled at the initial point locations to obtain track features $\mathbf{Q}$. Although features are derived from the full sequence, they are processed in overlapping sliding windows of $S=8$ frames, with a stride of $S/2$ (adjustable before training).

At each window step, CoTracker constructs a four-scale feature pyramid and cross-correlates it with $\mathbf{Q}$ to generate correlation maps. Sampling $7 \times 7$ pixel areas around the initial point locations produces correlation features $\mathbf{C}$. These, along with $\mathbf{Q}$, the point flows $\mathbf{P - P}_{t=0}$, their embeddings, and the visibility mask, form the transformer’s input. The input is both spatially and temporally encoded. The transformer architecture mirrors ViT-S \cite{dosovitskiy_image_2021} in layer count and dimensionality, but alternates between attention across frames and across point tracks. Cross-track attention enables efficient joint tracking. The transformer iteratively updates the coordinates and $\mathbf{Q}$, recalculating other components with each update. CoTracker typically performs four to six refinement iterations within each window before advancing to the next, repeating this process until the entire sequence is processed.

As a general-purpose tracking architecture, CoTracker has a large knowledge capacity and size, with over 24M parameters. Its computational demands are further increased by short sliding windows and iterative refinement because parts of CoTracker must run multiple times to process a sequence. The update to CoTracker2 increased the network size to 45M parameters by introducing virtual tracks in the transformer. Virtual tracks add cross-attention layers to reduce the total attention costs for large queries (e.g., tracking every single pixel), but they may be otherwise inefficient.

Previously, we proposed EchoTracker as a better fit for tracking in echocardiography, as it exploits domain-specific characteristics, such as limited motion between frames \cite{azad_echotracker_2024}. EchoTracker has a wider temporal context because it processes the whole sequence at once. Working in two stages, it first makes rough predictions from coarse features and then uses them to initialize second-stage tracking with higher-resolution features. It aggregates more information than CoTracker by using frame flow and point locations themselves as additional input to the location update module. Instead of a transformer, EchoTracker uses a series of eight 1D convolutional residual blocks to produce updates. This decision was motivated by the notion that transformers require large amounts of data. EchoTracker runs the location update module four times during the second stage for iterative refinement. The architecture has a lower parameter count (approximately 10M) than CoTracker and tends to execute much faster.

\subsection{Designing MyoTracker from CoTracker2}
To develop MyoTracker from CoTracker2 (Figure \ref{fig:myotracker}), we first removed all point visibility estimation components because point visibility was not annotated in the data. Point queries were limited to only the first frame, as later queries are unnecessary for downstream clinical tasks. We then modified the frame feature encoder (top-right in the figure) from a ResNet-18 variant to four simple double-convolution blocks with roughly a quarter of the original dimensionality in each layer. The search radius for tracking features was slightly reduced by shrinking the sampled area from $7\times7$ to $5\times5$ pixels, as the movement of the heart in the recordings was relatively constrained.

The transformer (bottom-left in the figure) for updating point locations by processing the aggregated features was simplified from six time-space attention blocks to four. The feature dimensionality was reduced to $d_{model}=64$ instead of $384$, with the underlying dense layers scaled from $1536$ to $64$ units. As the number of tracked points was relatively small in our case, the virtual track components in the transformer (a CoTracker2 feature) offered no benefits in the preliminary experiments and were discarded. The input to the transformer originally included temporal and spatial encoding. We completely removed the encoding, because the input features could feasibly contain sufficient information by themselves in the final design.

On the assumption that our transformer had sufficient complexity to estimate point displacement with a single pass, we dismantled the refinement loop to maximize speed and efficiency. The sliding windows were replaced with whole-sequence processing for the same reason and to widen the temporal context to overcome speckle decorrelation. As the network would now track all keypoint locations in the entire sequence at once without intermediary predictions, the initial point flow would always be zero. Therefore, the transformer input was modified to accept the embedded initial coordinates instead. This change allows the input to be spatially encoded by itself. The resulting architecture was relatively lightweight, with 0.32M parameters.

\subsection{Data Acquisition and Processing}
\begin{figure}
    \centering
    \includegraphics[width=\columnwidth]{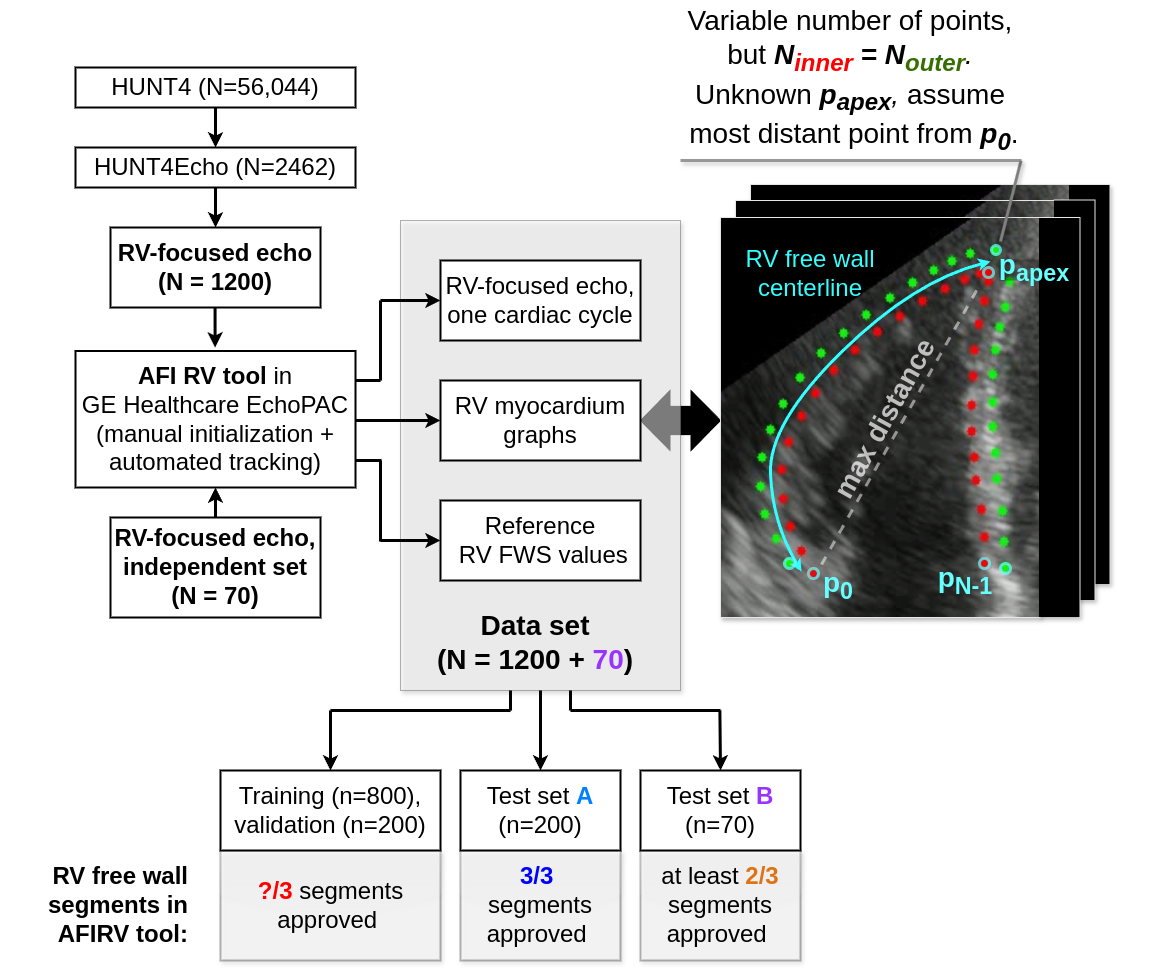}
    \caption{Data flow and processing steps. Processing RV-focused recordings with the AFI RV tool yields single-cycle recordings, RV myocardium keypoints, and corresponding RV FWS values. The software does not disclose the steps it takes to calculate RV FWS from the keypoints.}
    \label{fig:data_flow}
\end{figure}

Most of the data used in this work were obtained as part of the North-Trøndelag Health Study (HUNT4), conducted on informed, consenting volunteers in Trøndelag County, Norway \cite{asvold_cohort_2023}. In the HUNT4Echo sub-study, two operators performed echocardiography examinations on a subset (N=2462) of the volunteers using GE Healthcare Vivid E95 scanners. None of the participants had any significant heart disease \cite{nyberg_echocardiographic_2023}. Two senior cardiologists processed a subset of 1200 RV-focused recordings from this database with the AFI RV tool (a semi-automated RV analysis tool) in GE Healthcare EchoPAC (v204) software. The initial regions of interest were manually selected and adjusted by the annotators throughout the process until the software produced optimal tracking results. An independent set of 70 RV-focused recordings was also collected at St. Olav's University Hospital in Trondheim, Norway, to supplement the evaluation. This set was obtained from consenting volunteers as well, but the subjects were not guaranteed to be healthy. The data were processed in the same way as before, by an experienced clinician. The extracted one-cycle recordings, peak systolic RV FWS reference values, and keypoints for tracking comprised the full dataset, as illustrated in Figure \ref{fig:data_flow}.

Out of the 1200 HUNT4-sourced data samples, we reserved 800 for training the models, 200 for training validation, and 200 for testing (Test Set A). The 70 independent samples formed the second test set (Test Set B). Test Set A consisted only of the recordings where all three segments of the RV free wall were approved by the AFI RV tool in EchoPAC. For Test Set B, we allowed the recordings where at least two of the segments were approved. The recordings in Test Set A were 50-128 frames long (83$\pm$15), and the ones in Test Set B had 38-119 frames (77$\pm$20).

The extracted reference keypoint trajectories for the RV myocardium consisted of the inner and the outer boundary sub-graphs. The number of points $N$ in both sub-graphs was always equal, but it varied between recordings ($N \in [21,44]$). The first point $p_0$ in every sub-graph corresponded to the free wall (left) base point, and the last point $p_{N-1}$ denoted the septal (right) base point. The index of the apex point $p_{\text{apex}}$ was unknown, but we experimentally determined it as the point most distant to $p_0$. 

\subsection{Data augmentation and training setup}
The training data were heavily augmented through procedures affecting the video intensity values, geometry, as well as temporal resolution and direction. We started with some of the augmentations provided with the CoTracker package such as blur and regional blackout/replacement. Then, we applied rotation, zoom in/out, translation, random noise, frame skipping, and time reversal with a probability of 50\% each. Lastly, the order of points in the graph was always shuffled. To get fixed-size input for training, we oversampled the keypoint graphs to $N = 88$ points (maximum encountered in the data) by randomly duplicating the existing points. Likewise, we either randomly selected a chunk of frames or padded the video with its reverse to obtain fixed-length samples of $T = 64$ frames. The pixel intensity was constrained to $\left[0,1\right]$ through division by 255.

All three networks were trained on a single GeForce RTX 3090 GPU (24 GB video RAM) for 100\,000 steps, with validation performed at the end of each training epoch on the validation set. Only the model state from the best performing epoch in validation was retained. Setting up CoTracker2, we changed the model input resolution from 384$\times$512 to 256$\times$256 but reduced the encoder stride from 8 to 4, thus maintaining comparable resolution in the extracted features. The sliding window length remained at $S = 8$ frames, while the number of virtual tracks was set to $N_{\text{virt}}=16$ (maintaining $N_\text{virt} << N = 88$). We trained CoTracker2 with its default loss (a variant of mean absolute error applied to point coordinates) and default optimizer (AdamW, a linear warm-up to the peak learning rate of $5\cdot10^{-4}$ and linear decay). The same held for EchoTracker, which used similar settings. We could only afford a batch size of two for CoTracker2 and EchoTracker due to their GPU memory requirements. For MyoTracker, we used a batch size of 8 and a learning rate of $10^{-3}\cdot0.99995^{\textit{step}}$ (exponential decay) without warm-up. MyoTracker's loss function used the absolute errors of point coordinates and those of point flows between the frames:
\begin{equation}
    \mathcal{L}(\mathbf{P}, \hat{\mathbf{P}}) = \lvert\mathbf{P}-\hat{\mathbf{P}}\rvert + \left\lvert \frac{\partial\mathbf{P}}{\partial t} - \frac{\partial\hat{\mathbf{P}}}{\partial t}\right\rvert
\end{equation}
where $\mathbf{P}$ and $\hat{\mathbf{P}}$ are the reference and the predicted point coordinates, respectively. Inclusion of point flows was intended to teach the network smooth motion patterns in the absence of any refinement procedures.

\subsection{Strain calculation algorithm}
Since EchoPAC relies on undisclosed algorithms to calculate clinical metrics from tracking results, we implemented our own method to compute RV strain. The most commonly used metric for strain measurements in the RV is the Free Wall Strain (RV FWS). Similar to the definition of LV GLS, RV FWS is calculated as the ratio between the free wall shortening compared to its total length at end-diastole (ED). At a given time \textit{t},
\begin{equation}
   \text{RV FWS}(t) = \frac{|\text{FW}_{t}| - |\text{FW}_{\text{ED}}|}{|\text{FW}_{\text{ED}}|} \cdot 100\%
\end{equation}
where $\text{FW}_t$ is the centerline of the free wall at time $t$. We extracted the centerline by taking the mean between the inner and the outer sub-graphs from the first point ($p_0$, left base) up to the apex point ($p_{\text{apex}}$, geometrically the furthest from $p_0$). The peak systolic RV FWS values were assumed to be at the time points when the entire RV myocardium was at its minimum length.

\renewcommand{\arraystretch}{1.25}
\begin{table*}
\caption{Model performance characteristics on an RTX 3090 24GB GPU (PyTorch 2.3.1, CUDA 12.1).\\ The values are reported for 100 points tracked in a 100-frame sequence.}
\label{table_performance}
\centering
\resizebox{\textwidth}{!}{%
\begin{tabular}{ccccc}
    \hline
    & \textbf{Parameters (M)} & \textbf{Model File (MB)} & \textbf{GPU Memory (MB)*} & \textbf{Inference Time (ms)} \\ \hline
    MyoTracker & 0.317& 1.3& 958 & 23\\
    CoTracker2 \cite{karaev_cotracker_2023} & 45.434& 189.8& 2928 & 1692**\\
    EchoTracker \cite{azad_echotracker_2024} & 10.171& 40.8 & 5935 & 244\\ \hline
    \multicolumn{5}{l}{\rule{0pt}{4ex}*From \texttt{torch.cuda.memory\_summary()} during inference. Excludes CUDA context (300-500MB).}\\
    \multicolumn{5}{l}{**Scales with the number of refinement iterations and the number of sliding window steps to cover the video.}
\end{tabular}
}
\end{table*}
\renewcommand{\arraystretch}{1.0}

\renewcommand{\arraystretch}{1.25}
\begin{table*}
\caption{Tracking accuracy and drift of the trained models (mean $\pm$ standard deviation between sequences).\\ Pixel distances are listed for downsampled sequences (256$\times$256 pixel resolution).}
\label{table_keypoint_accuracy}
\centering
\resizebox{\textwidth}{!}{%
\begin{tabular}{rccccccccccccc}
    \hline
     & \multicolumn{3}{c}{\textbf{Average Trajectory Error}$\downarrow$} &  &  & \multicolumn{3}{c}{\textbf{End Trajectory Error}$\downarrow$} &  &  & \multicolumn{3}{c}{\textbf{ED-to-ED Drift}$\downarrow$} \\ \hline
     & \textit{pixels} &  & \textit{mm} &  &  & \textit{pixels} &  & \textit{mm} &  &  & \textit{pixels} &  & \textit{mm} \\ \cline{2-2} \cline{4-4} \cline{7-7} \cline{9-9} \cline{12-12} \cline{14-14} 
    \multicolumn{1}{l}{\textbf{Test Set A:}} &  &  &  &  &  &  &  &  &  &  &  &  &  \\
    MyoTracker  & \textbf{3.84}$\pm$\textbf{1.00} &  & \textbf{2.00}$\pm$\textbf{0.53} &  &  & \textbf{5.57}$\pm$\textbf{1.98} &  & \textbf{2.92}$\pm$\textbf{1.08} &  &  & \textbf{5.91}$\pm$\textbf{4.37} &  & \textbf{2.91}$\pm$\textbf{2.14} \\
    CoTracker2 \cite{karaev_cotracker_2023} & 4.35$\pm$1.31 &  & 2.25$\pm$0.63 &  &  & 5.85$\pm$1.97 &  & 3.06$\pm$1.04 &  &  & 6.66$\pm$3.77 &  & 3.31$\pm$1.87 \\
    EchoTracker \cite{azad_echotracker_2024} & 4.84$\pm$1.38 &  & 2.52$\pm$0.72 &  &  & 6.76$\pm$2.41 &  & 3.55$\pm$1.28 &  &  & 6.53$\pm$3.79 &  & 3.27$\pm$2.06 \\
     &  &  &  &  &  &  &  &  &  &  &  &  &  \\
    
    \multicolumn{1}{l}{\textbf{Test Set B:}} &  &  &  &  &  &  &  &  &  &  &  &  \\
    MyoTracker  & \textbf{3.36}$\pm$\textbf{1.04} &  & \textbf{1.94}$\pm$\textbf{0.57} &  &  & \textbf{4.35}$\pm$\textbf{2.09} &  & \textbf{2.47}$\pm$\textbf{1.05} &  &  & \textbf{3.86}$\pm$\textbf{2.83} &  & \textbf{2.14}$\pm$\textbf{1.51} \\ 
    CoTracker2 \cite{karaev_cotracker_2023} & 3.76$\pm$1.23 &  & 2.16$\pm$0.60 &  &  & 4.71$\pm$2.38 &  & 2.68$\pm$1.15 &  &  & 4.63$\pm$2.21 &  & 2.59$\pm$1.11 \\
    EchoTracker \cite{azad_echotracker_2024} & 4.33$\pm$1.20 &  & 2.50$\pm$0.63&  &  & 5.53$\pm$2.48 &  & 3.16$\pm$1.24 &  &  & 4.06$\pm$2.21 &  & 2.30$\pm$1.18 \\ \hline
    \multicolumn{14}{l}{\rule{0pt}{4ex}\textbf{Average Trajectory Error:} average of all keypoint errors in \textit{all frames} of the sequence.}\\
    \multicolumn{14}{l}{\textbf{End Trajectory Error:} average of all keypoint errors in the \textit{last frame} of the sequence (assumed as the second end-diastole).}\\
    \multicolumn{14}{l}{\textbf{ED-to-ED Drift:} distance between the first and the last frame (assumed as the first and the second end-diastole) keypoints.}
    \end{tabular}%
}
\end{table*}
\renewcommand{\arraystretch}{1.0}

\section{Experiments and Results}
We compared three point tracking models (MyoTracker, CoTracker2, and EchoTracker), first assessing their complexity and inference speed. The models were then applied to test sets A and B in order to evaluate the point tracking accuracy. Lastly, we used the RV FWS calculation algorithm with the predicted points from each model and compared the results to the reference values. As an addition, we conducted an ablation study on MyoTracker, wherein various components from CoTracker2 were swapped back into the architecture to measure their influence on model complexity, speed, tracking accuracy, and RV FWS estimation accuracy.

\subsection{Model Characteristics}
The trained models’ characteristics in terms of size, GPU memory consumption, and inference speed were recorded and are listed in Table \ref{table_performance}. Each model was kept in its original PyTorch format, with the default 32-bit precision, and without any additional inference optimizations. CoTracker2 was run with six refinement iterations of the transformer, as prescribed by the authors, while EchoTracker was run with four iterations. Measurements were performed using mock sequences of 100 frames with 100 tracked points. Inference time was calculated as the average of 100 sequence predictions following a warm-up phase of 100 predictions. Under this setup, MyoTracker was the fastest, while occupying the least disk space and consuming the least GPU memory.

\subsection{Tracking Accuracy}
\begin{figure*}
    \centering
    \includegraphics[width=\textwidth]{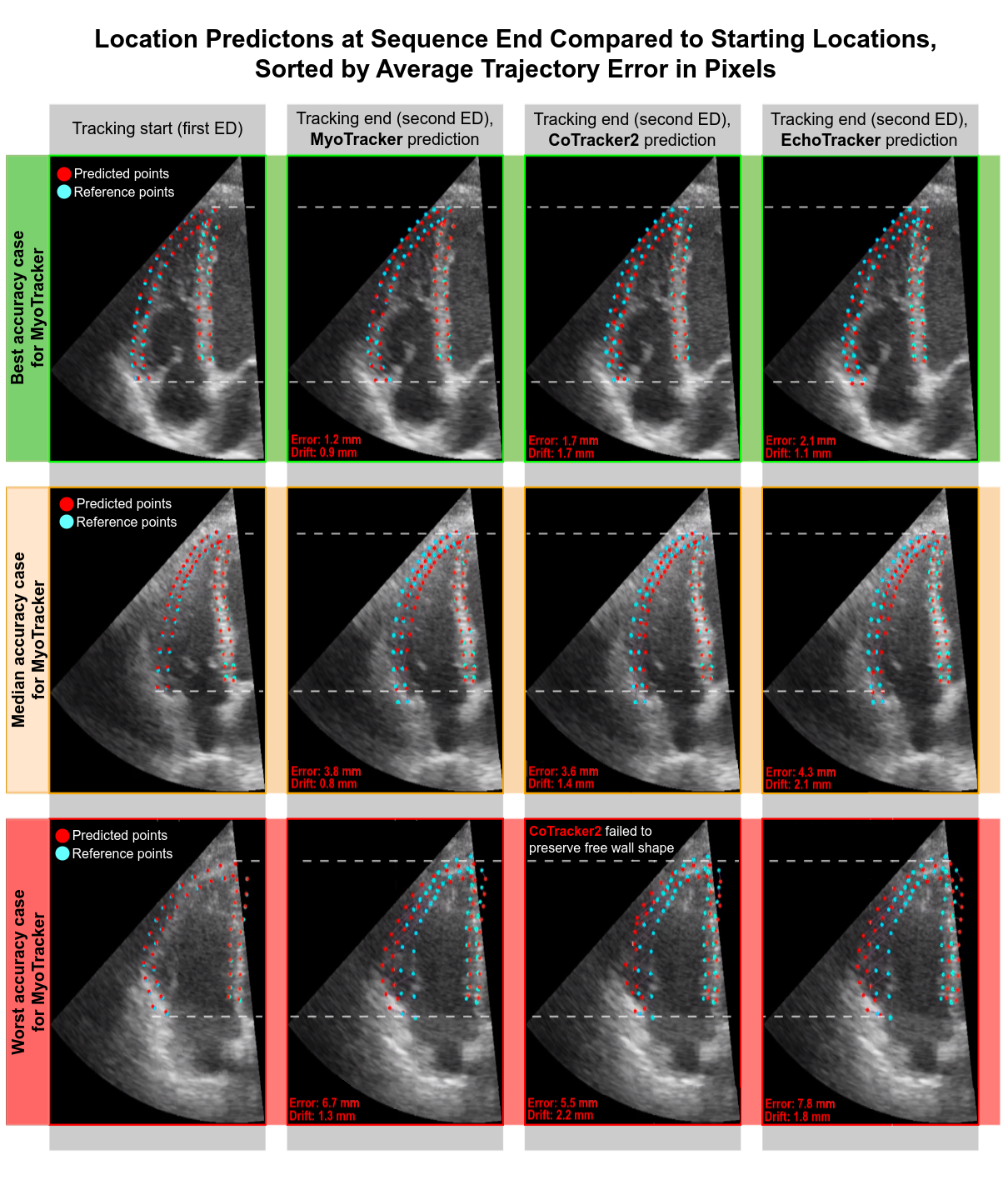}
    \caption{A qualitative look at the models' predictions. The left column shows sequence frames close to the start (first end-diastole), and the other columns display the last frame (second end-diastole) predictions for the same sequence using each of the models. The cases are sorted by the average trajectory error (in pixels) in MyoTracker predictions. End trajectory error and drift are provided for each prediction. Support lines are drawn at equivalent locations between the frames to illustrate potential drift after a full cardiac cycle.}
    \label{fig:visuals}
\end{figure*}

Using the keypoints generated by the AFI RV tool in EchoPAC as the reference, we measured the point tracking accuracy for each trained model on both datasets and recorded the results in Table \ref{table_keypoint_accuracy}. The best, median, and worst cases for MyoTracker are separately displayed in Figure \ref{fig:visuals}. We selected the average trajectory error (averaged across all frames) and the end trajectory error (at the final frame) as our evaluation metrics. The point drift between the first and last frames, both assumed to correspond to end-diastole, was also considered. All metrics were reported in pixels (at 256$\times$256 pixel video resolution) and in millimeters.

\begin{figure*}
    \centering
    \includegraphics[height=0.9\textheight]{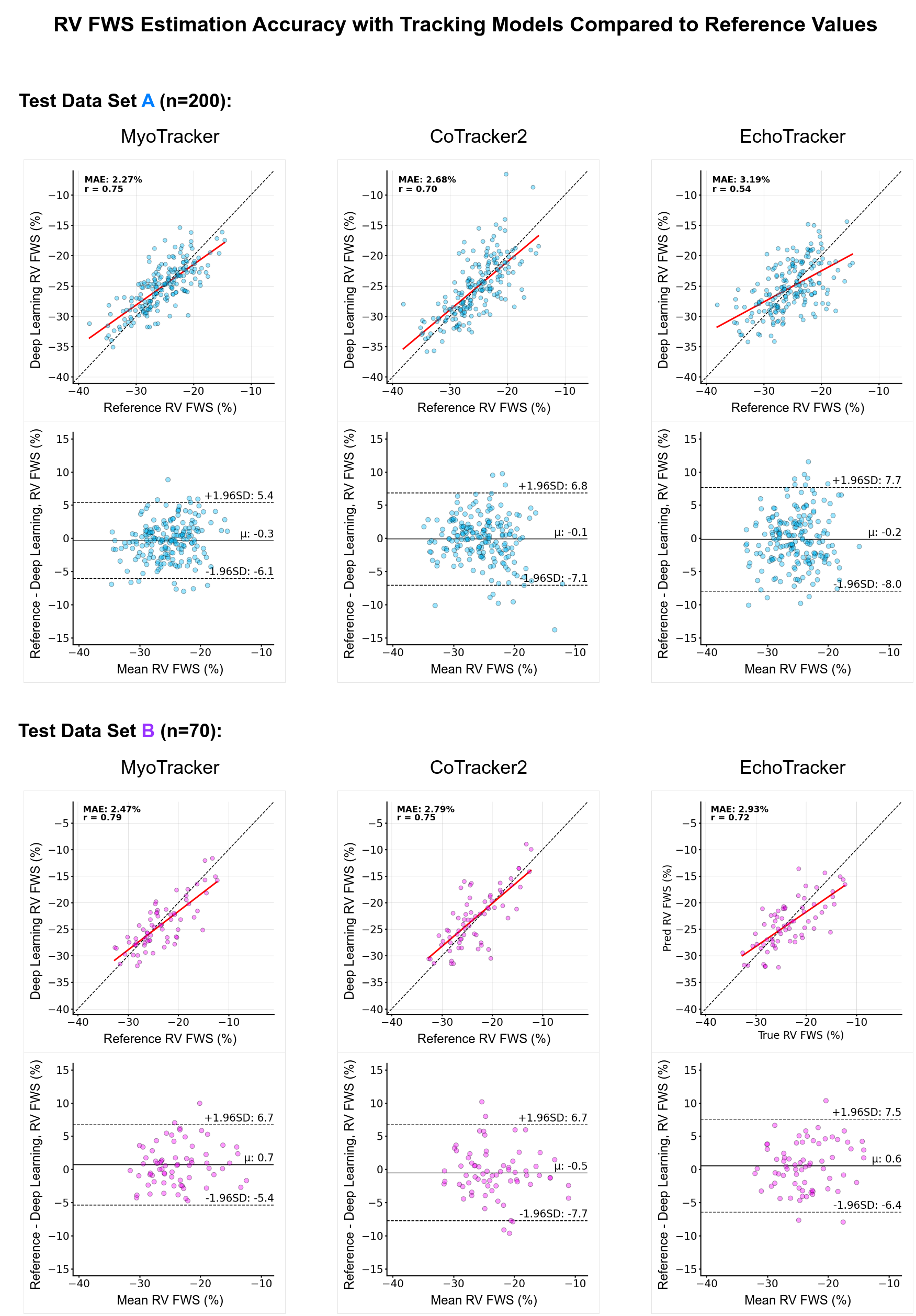}
    \caption{Peak systolic RV FWS estimation accuracy with the trained models and our strain calculation algorithm when compared to the reference values. The differences in both tracking and strain calculation methods lead to compounding errors.}
    \label{fig:strain_accuracy}
\end{figure*}

\renewcommand{\arraystretch}{1.25}
\begin{table*}
\caption{MyoTracker ablation study: reverting a given MyoTracker component back to CoTracker2 equivalent.\\ Values in brackets indicate relative change from baseline MyoTracker. Lower values are better.}
\label{table_ablation}
\centering
\resizebox{\textwidth}{!}{%
\begin{tabular}{rccccccccc}
    \hline
    & \textbf{Parameters} &  & \textbf{GPU Memory$\downarrow$} &  & \textbf{Inference Time$\downarrow$} &  & \textbf{Avg. Traj. Error$\downarrow$} &  & \textbf{RV FWS Abs. Error$\downarrow$} \\ \hline
     & \textit{M} &  & \textit{MB} &  & \textit{ms/sequence} &  & \textit{pixels, in Test Set A} &  & \textit{\%, in Test Set A} \\ \cline{2-2} \cline{4-4} \cline{6-6} \cline{8-8} \cline{10-10} 
    Baseline MyoTracker & \textbf{0.317} &  & 958 &  & \textbf{23} &  & 3.84$\pm$1.00 &  & 2.27$\pm$1.87 \\[0.1cm]
    \multicolumn{1}{l}{\textbf{Baseline + component:}} &  &  &  &  & &  & & & \\
    + lower batch size (B=2) & 0.317 (+0\%) &  & 958 (+0\%) &  & 23 (+0\%) &  & 3.91$\pm$1.03 (\textcolor{red}{+2\%}) &  & 2.46$\pm$2.09 (\textcolor{red}{+8\%}) \\
    + ResNet-18 encoder (d=128) & 2.845 (\textcolor{red}{+798\%}) &  & 1908 (\textcolor{red}{\textbf{+99\%}}) &  & 79 (\textcolor{red}{+243\%}) &  & 3.80$\pm$1.05 (\textcolor{darkgreen}{-1\%}) &  & 2.26$\pm$1.81 (+0\%) \\
    + wider sampling kernel (7x7) & 0.323 (\textcolor{red}{+2\%}) &  & 956 (+0\%) &  & 23 (+0\%) &  & 3.85$\pm$1.02 (+0\%) &  & 2.31$\pm$1.87 (\textcolor{red}{+2\%}) \\
    + spatial/temporal encoding & 0.317 (+0\%) &  & 979 (\textcolor{red}{+2\%}) &  & 23 (+0\%) &  & 4.02$\pm$1.06 (\textcolor{red}{+5\%}) &  & 2.48$\pm$2.14 (\textcolor{red}{+9\%}) \\
    + 6-layer transformer (d=384) & 21.454 (\textcolor{red}{\textbf{+6668\%}})&  & 803 (\textcolor{darkgreen}{-16\%}) &  & 31 (\textcolor{red}{+35\%}) &  & 4.22$\pm$1.17 (\textcolor{red}{+10\%}) &  & 2.48$\pm$2.10 (\textcolor{red}{+9\%}) \\
    + virtual tracks (N=16) & 0.520 (\textcolor{red}{+64\%}) &  & 960 (+0\%) &  & 25 (\textcolor{red}{+9\%}) &  & 3.95$\pm$1.11 (\textcolor{red}{+3\%}) &  & 2.41$\pm$2.00 (\textcolor{red}{+6\%}) \\
    + iterative refinement (x6) & 0.325 (\textcolor{red}{+3\%}) &  & 860 (\textcolor{darkgreen}{-10\%}) &  & 46 (\textcolor{red}{+100\%}) &  & \textbf{3.65}$\pm$\textbf{1.04} (\textcolor{darkgreen}{\textbf{-5\%}}) &  & \textbf{2.17}$\pm$\textbf{1.90} (\textcolor{darkgreen}{\textbf{-4\%}}) \\
    + sliding windows (S=8) & 0.321 (\textcolor{red}{+1\%}) &  & 
    \textbf{573} (\textcolor{darkgreen}{\textbf{-40\%}}) &  & 136 (\textcolor{red}{\textbf{+491\%}}) &  & 4.52$\pm$1.21 (\textcolor{red}{\textbf{+18\%}}) &  & 2.57$\pm$2.08 (\textcolor{red}{\textbf{+13\%}}) \\ \hline
    \multicolumn{10}{l}{\rule{0pt}{4ex}\textbf{GPU Memory} and \textbf{Inference Time} were measured on dummy sequences with 100 keypoints tracked through 100 frames.}\\
    \multicolumn{10}{l}{Loss function changed to support multiple predictions when re-enabling \textbf{iterative refinement} and \textbf{sliding windows}.}\\
    \multicolumn{10}{l}{Numerical values in boldface indicate the best result among the models.}\\
\end{tabular}%
}
\end{table*}
\renewcommand{\arraystretch}{1.0}

\subsection{Strain Estimation Accuracy}
We used the keypoints predicted by each trained model to calculate the RV FWS values with our algorithm and compared the results to the reference RV FWS (Figure \ref{fig:strain_accuracy}).

\subsection{MyoTracker Ablation Study}
We altered the MyoTracker architecture by changing certain components back to their CoTracker2 counterparts, one at a time, and trained the resulting models. The experiments included lowering the batch size to two (same as in CoTracker2 and EchoTracker training), using the original frame encoder, restoring the larger six-layer transformer, enabling the refinement loop, processing the sequences in sliding windows, and more. For each experiment, we recorded the complexity, speed, and accuracy of the resulting model (Table \ref{table_ablation}).

\section{Discussion}
We explored the use of several deep learning algorithms for point tracking in 2D echocardiography recordings of the RV myocardium and proposed one of our own. In our previous work on RV segmentation and keypoint detection, we hypothesized that echocardiogram interpretation tasks can be performed with less complex deep learning models than those typically proposed in the literature \cite{chernyshov_automated_2024}. We designed MyoTracker based on this notion. The architecture compared favorably on all accuracy metrics against both the baseline CoTracker2 and the recently proposed EchoTracker. When using the predicted point locations to calculate the peak systolic RV FWS values, our model showed better agreement with the reference values than the other two models. It also maintained superior performance in other areas, achieving a higher inference speed and lower GPU utilization compared to the other networks. In a clinical device or software, efficient models could improve the workflow by producing measurements faster after acquisition or by enabling additional functions to run in parallel using available memory. Other possibilities include deployment on mobile devices and real-time measurements during scanning.

The subsequent ablation study (Table \ref{table_ablation}) demonstrated that MyoTracker benefited from most of its design choices. Allowing the network to access the entire sequence at once had the most significant positive impact on accuracy and shape-preserving properties. Limiting the temporal context to $S = 8$ frames resulted in an 18\% higher (worse) keypoint error on Test Set A. Reducing the overall network size allowed for larger batch sizes during training, which improved training stability and provided minor performance gains. The lower complexity of both the frame encoder and transformer remained sufficient for the task. One design choice that produced mixed results was the removal of iterative refinement. Reintroducing it led to slight accuracy improvements but increased runtime by approximately 20\% of the baseline for each additional transformer pass. This trade-off may be acceptable depending on the application, but the refinement loop can complicate model serialization and hinder interoperability with frameworks other than PyTorch. Interestingly, configurations with refinement iterations and larger transformer sizes showed lower GPU memory usage compared to the baseline, which may be due to specifics in PyTorch's optimization processes.

Assessment of individual predictions from MyoTracker and the compared models showed that trajectory error did not always capture all the desired properties as either a loss function or a primary accuracy metric. For example, CoTracker2 occasionally achieved better keypoint accuracy on specific recordings (Figure \ref{fig:visuals}, bottom row) but introduced distortions to the RV free wall. In contrast, MyoTracker and EchoTracker preserved myocardial shape and maintained consistent inner-outer boundary distances, likely because both models processed the entire sequence within their temporal context. Notably, the cases in which MyoTracker showed the greatest end trajectory error often coincided with significant divergence in the reference points toward the end of the sequence.

Using MyoTracker’s tracking results to calculate RV FWS with our algorithm yielded a bias of -0.3\% and 95\% limits of agreement (LoA) ranging from -6.1\% to 5.4\% compared to the reference values in Test Set A (Figure \ref{fig:strain_accuracy}). This performance is comparable to the previously reported interobserver variability in the HUNT4 dataset, which showed a bias of -0.2\% (95\% LoA: -6.5\% to 6.2\%) \cite{nyberg_echocardiographic_2023}. However, MyoTracker still struggled with extreme RV FWS values, tending to underestimate higher values and overestimate lower ones. EchoTracker performed worse in this regard, particularly on Test Set A, which contained a wider range of RV FWS values and more dynamic heart motion than Test Set B. EchoTracker was originally designed and validated for LV tracking, whereas RV GLS and especially RV FWS tend to have greater magnitudes than LV GLS \cite{morris_normal_2017, wang_defining_2021, nyberg_echocardiographic_2023}. CoTracker2’s performance generally fell between that of MyoTracker and EchoTracker, though the Bland-Altman analysis indicated that it produced some of the most extreme outliers.

Several limitations emerged during this study, primarily related to the quality and characteristics of the data. RV-focused echocardiograms were challenging to analyze because the imaging views are poorly standardized, and the image quality is often suboptimal. The RV free wall, which was of the greatest interest, tended to be the least clearly visible region, with frequent dropouts and out-of-plane motion.

Given the significant resources required to manually annotate RV myocardium motion in a large number of samples, we deemed the collection of fully manual labels unfeasible. Using commercial clinical software allowed us to acquire a sizable dataset quickly but introduced uncertainty in the ground truth. The point tracking quality provided by the software was not entirely reliable, which may have caused our models to learn undesirable behaviors. Furthermore, the vendor’s strain calculation methods were undisclosed, necessitating the development of our own algorithm. Potential differences between these algorithms likely affected RV FWS estimation accuracy. These factors complicated our evaluation procedures and limited the performance ceiling for our models given the available data. At the same time, the imperfections in commercial tools highlighted the need for this and similar research.

Lastly, the lack of optical flow data for the RV limited our ability to directly compare MyoTracker to state-of-the-art optical flow estimation architectures in echocardiography \cite{ostvik_myocardial_2021, evain_motion_2022}. In previous work on EchoTracker for the LV myocardium specifically \cite{azad_echotracker_2024}, data were available in both keypoint and optical flow formats. EchoTracker achieved greater LV GLS estimation accuracy compared to optical flow methods in that context \cite{salte_artificial_2021, salte_deep_2023}. In this study, MyoTracker slightly outperformed EchoTracker, suggesting that it could deliver performance equivalent to or better than optical flow-based approaches. However, further evaluation of MyoTracker using LV myocardial data will be necessary to confirm this potential.

\section{Conclusion}
We developed MyoTracker, a compact deep learning architecture for myocardial tracking in 2D echocardiography recordings. It integrates key components from existing architectures with extended temporal context support. MyoTracker outperforms evaluated alternatives in point tracking accuracy on RV-focused recordings while offering significantly greater computational efficiency and speed through optimized component design. RV FWS values estimated from its predictions show strong correlation ($r \geq 0.75$) with reference values from commercial software. In clinical workflows, MyoTracker could greatly accelerate RV measurements and is well-suited for deployment on mobile devices.

\section*{Acknowledgment}
This work was funded by the Research Council of Norway through the ProCardio Center for Innovation (project number 309762) and the Center for Innovative Ultrasound Solutions (project number 237887). GE Healthcare is a partner in both projects.

\bibliographystyle{ieeetr}


\begin{thebibliography}{34}

\bibitem{koopman_comparison_2010}
L.~P. Koopman, C.~Slorach, W.~Hui, C.~Manlhiot, B.~W. McCrindle, M.~K. Friedberg, E.~T. Jaeggi, and L.~Mertens, ``Comparison between {Different} {Speckle} {Tracking} and {Color} {Tissue} {Doppler} {Techniques} to {Measure} {Global} and {Regional} {Myocardial} {Deformation} in {Children},'' {\em Journal of the American Society of Echocardiography}, vol.~23, pp.~919--928, Sept. 2010.

\bibitem{leitman_two-dimensional_2004}
M.~Leitman, P.~Lysyansky, S.~Sidenko, V.~Shir, E.~Peleg, M.~Binenbaum, E.~Kaluski, R.~Krakover, and Z.~Vered, ``Two-dimensional strain–a novel software for real-time quantitative echocardiographic assessment of myocardial function,'' {\em Journal of the American Society of Echocardiography}, vol.~17, pp.~1021--1029, Oct. 2004.

\bibitem{heyde_elastic_2013}
B.~Heyde, R.~Jasaityte, D.~Barbosa, V.~Robesyn, S.~Bouchez, P.~Wouters, F.~Maes, P.~Claus, and J.~D'hooge, ``Elastic {Image} {Registration} {Versus} {Speckle} {Tracking} for 2-{D} {Myocardial} {Motion} {Estimation}: {A} {Direct} {Comparison} {In} {Vivo},'' {\em IEEE Transactions on Medical Imaging}, vol.~32, pp.~449--459, Feb. 2013.

\bibitem{alessandrini_myocardial_2013}
M.~Alessandrini, A.~Basarab, H.~Liebgott, and O.~Bernard, ``Myocardial {Motion} {Estimation} {From} {Medical} {Images} {Using} the {Monogenic} {Signal},'' {\em IEEE Transactions on Image Processing}, vol.~22, pp.~1084--1095, Mar. 2013.

\bibitem{sitia_speckle_2010}
S.~Sitia, L.~Tomasoni, and M.~Turiel, ``Speckle tracking echocardiography: {A} new approach to myocardial function,'' {\em World Journal of Cardiology}, vol.~2, pp.~1--5, Jan. 2010.

\bibitem{dhooge_two-dimensional_2016}
J.~D'hooge, D.~Barbosa, H.~Gao, P.~Claus, D.~Prater, J.~Hamilton, P.~Lysyansky, Y.~Abe, Y.~Ito, H.~Houle, S.~Pedri, R.~Baumann, J.~Thomas, L.~P. Badano, and {on behalf of the EACVI/ASE/Industry Task Force to Standardize Deformation Imaging}, ``Two-dimensional speckle tracking echocardiography: standardization efforts based on synthetic ultrasound data,'' {\em European Heart Journal - Cardiovascular Imaging}, vol.~17, pp.~693--701, June 2016.

\bibitem{dosovitskiy_flownet_2015}
A.~Dosovitskiy, P.~Fischer, E.~Ilg, P.~Häusser, C.~Hazirbas, V.~Golkov, P.~v.~d. Smagt, D.~Cremers, and T.~Brox, ``{FlowNet}: {Learning} {Optical} {Flow} with {Convolutional} {Networks},'' in {\em 2015 {IEEE} {International} {Conference} on {Computer} {Vision} ({ICCV})}, pp.~2758--2766, Dec. 2015.
\newblock ISSN: 2380-7504.

\bibitem{sun_pwc-net_2018}
D.~Sun, X.~Yang, M.-Y. Liu, and J.~Kautz, ``{PWC}-{Net}: {CNNs} for {Optical} {Flow} {Using} {Pyramid}, {Warping}, and {Cost} {Volume},'' in {\em 2018 {IEEE}/{CVF} {Conference} on {Computer} {Vision} and {Pattern} {Recognition}}, pp.~8934--8943, June 2018.
\newblock ISSN: 2575-7075.

\bibitem{huang_flowformer_2022}
Z.~Huang, X.~Shi, C.~Zhang, Q.~Wang, K.~C. Cheung, H.~Qin, J.~Dai, and H.~Li, ``{FlowFormer}: {A} {Transformer} {Architecture} for {Optical} {Flow},'' Sept. 2022.
\newblock arXiv:2203.16194.

\bibitem{luo_flowdiffuser_2024}
A.~Luo, X.~Li, F.~Yang, J.~Liu, H.~Fan, and S.~Liu, ``{FlowDiffuser}: {Advancing} {Optical} {Flow} {Estimation} with {Diffusion} {Models},'' in {\em 2024 {IEEE}/{CVF} {Conference} on {Computer} {Vision} and {Pattern} {Recognition} ({CVPR})}, (Seattle, WA, USA), pp.~19167--19176, IEEE, June 2024.

\bibitem{doersch_tap-vid_2023}
C.~Doersch, A.~Gupta, L.~Markeeva, A.~Recasens, L.~Smaira, Y.~Aytar, J.~Carreira, A.~Zisserman, and Y.~Yang, ``{TAP}-{Vid}: {A} {Benchmark} for {Tracking} {Any} {Point} in a {Video},'' Mar. 2023.
\newblock Issue: arXiv:2211.03726 arXiv:2211.03726 [cs, stat].

\bibitem{harley_particle_2022}
A.~W. Harley, Z.~Fang, and K.~Fragkiadaki, ``Particle {Video} {Revisited}: {Tracking} {Through} {Occlusions} {Using} {Point} {Trajectories},'' July 2022.
\newblock Issue: arXiv:2204.04153 arXiv:2204.04153 [cs].

\bibitem{zheng_pointodyssey_2023}
Y.~Zheng, A.~W. Harley, B.~Shen, G.~Wetzstein, and L.~J. Guibas, ``{PointOdyssey}: {A} {Large}-{Scale} {Synthetic} {Dataset} for {Long}-{Term} {Point} {Tracking},'' July 2023.
\newblock Issue: arXiv:2307.15055 arXiv:2307.15055 [cs].

\bibitem{doersch_tapir_2023}
C.~Doersch, Y.~Yang, M.~Vecerik, D.~Gokay, A.~Gupta, Y.~Aytar, J.~Carreira, and A.~Zisserman, ``{TAPIR}: {Tracking} {Any} {Point} with per-frame {Initialization} and temporal {Refinement},'' Aug. 2023.
\newblock Issue: arXiv:2306.08637 arXiv:2306.08637 [cs].

\bibitem{karaev_cotracker_2023}
N.~Karaev, I.~Rocco, B.~Graham, N.~Neverova, A.~Vedaldi, and C.~Rupprecht, ``{CoTracker}: {It} is {Better} to {Track} {Together},'' Dec. 2023.
\newblock Issue: arXiv:2307.07635 arXiv:2307.07635 [cs].

\bibitem{ostvik_myocardial_2021}
A.~Østvik, I.~M. Salte, E.~Smistad, T.~M. Nguyen, D.~Melichova, H.~Brunvand, K.~Haugaa, T.~Edvardsen, B.~Grenne, and L.~Løvstakken, ``Myocardial {Function} {Imaging} in {Echocardiography} {Using} {Deep} {Learning},'' {\em IEEE Transactions on Medical Imaging}, vol.~40, no.~5, pp.~1340--1351, 2021.

\bibitem{evain_motion_2022}
E.~Evain, Y.~Sun, K.~Faraz, D.~Garcia, E.~Saloux, B.~L. Gerber, M.~De~Craene, and O.~Bernard, ``Motion {Estimation} by {Deep} {Learning} in {2D} {Echocardiography}: {Synthetic} {Dataset} and {Validation},'' {\em IEEE Transactions on Medical Imaging}, vol.~41, pp.~1911--1924, Aug. 2022.

\bibitem{azad_echotracker_2024}
M.~A. Azad, A.~Chernyshov, J.~Nyberg, I.~Tveten, L.~Lovstakken, H.~Dalen, B.~Grenne, and A.~Østvik, ``{EchoTracker}: {Advancing} {Myocardial} {Point} {Tracking} in {Echocardiography},'' in {\em Medical {Image} {Computing} and {Computer} {Assisted} {Intervention} – {MICCAI} 2024}, pp.~645--655, Springer Nature Switzerland, 2024.

\bibitem{ouyang_video-based_2020}
D.~Ouyang, B.~He, A.~Ghorbani, N.~Yuan, J.~Ebinger, C.~P. Langlotz, P.~A. Heidenreich, R.~A. Harrington, D.~H. Liang, E.~A. Ashley, and J.~Y. Zou, ``Video-based {AI} for beat-to-beat assessment of cardiac function,'' {\em Nature}, vol.~580, pp.~252--256, Apr. 2020.

\bibitem{wei_temporal-consistent_2020}
H.~Wei, H.~Cao, Y.~Cao, Y.~Zhou, W.~Xue, D.~Ni, and S.~Li, ``Temporal-{Consistent} {Segmentation} of {Echocardiography} with {Co}-learning from {Appearance} and {Shape},'' in {\em Medical {Image} {Computing} and {Computer} {Assisted} {Intervention} – {MICCAI} 2020} (A.~L. Martel, P.~Abolmaesumi, D.~Stoyanov, D.~Mateus, M.~A. Zuluaga, S.~K. Zhou, D.~Racoceanu, and L.~Joskowicz, eds.), (Cham), pp.~623--632, Springer International Publishing, 2020.

\bibitem{painchaud_echocardiography_2022}
N.~Painchaud, N.~Duchateau, O.~Bernard, and P.-M. Jodoin, ``Echocardiography {Segmentation} with {Enforced} {Temporal} {Consistency},'' May 2022.
\newblock arXiv:2112.02102.

\bibitem{deng_myocardial_2022}
Y.~Deng, P.~Cai, L.~Zhang, X.~Cao, Y.~Chen, S.~Jiang, Z.~Zhuang, and B.~Wang, ``Myocardial strain analysis of echocardiography based on deep learning,'' {\em Frontiers in Cardiovascular Medicine}, vol.~9, p.~1067760, Dec. 2022.

\bibitem{hu_automated_2024}
J.~Hu, S.~H. Olaisen, E.~Smistad, H.~Dalen, and L.~Lovstakken, ``Automated 2-{D} and 3-{D} {Left} {Atrial} {Volume} {Measurements} {Using} {Deep} {Learning},'' {\em Ultrasound in Medicine \& Biology}, vol.~50, pp.~47--56, Jan. 2024.

\bibitem{chernyshov_automated_2024}
A.~Chernyshov, J.~F. Grue, J.~Nyberg, B.~Grenne, H.~Dalen, S.~A. Aase, A.~Østvik, and L.~Lovstakken, ``Automated {Segmentation} and {Quantification} of the {Right} {Ventricle} in 2-{D} {Echocardiography},'' {\em Ultrasound in Medicine \& Biology}, vol.~50, pp.~540--548, Apr. 2024.

\bibitem{rudski_guidelines_2010}
L.~G. Rudski, W.~W. Lai, J.~Afilalo, L.~Hua, M.~D. Handschumacher, K.~Chandrasekaran, S.~D. Solomon, E.~K. Louie, and N.~B. Schiller, ``Guidelines for the {Echocardiographic} {Assessment} of the {Right} {Heart} in {Adults}: {A} {Report} from the {American} {Society} of {Echocardiography}: {Endorsed} by the {European} {Association} of {Echocardiography}, a registered branch of the {European} {Society} of {Cardiology}, and the {Canadian} {Society} of {Echocardiography},'' {\em Journal of the American Society of Echocardiography}, vol.~23, pp.~685--713, July 2010.

\bibitem{kossaify_echocardiographic_2015}
A.~Kossaify, ``Echocardiographic {Assessment} of the {Right} {Ventricle}, from the {Conventional} {Approach} to {Speckle} {Tracking} and {Three}-{Dimensional} {Imaging}, and {Insights} into the “{Right} {Way}” to {Explore} the {Forgotten} {Chamber},'' {\em Clinical Medicine Insights. Cardiology}, vol.~9, pp.~65--75, July 2015.

\bibitem{mcerlane_feasibility_2023}
J.~McErlane, B.~Shelley, and P.~McCall, ``Feasibility of 2-dimensional speckle tracking echocardiography strain analysis of the right ventricle with trans-thoracic echocardiography in intensive care: a literature review and meta-analysis,'' {\em Echo Research \& Practice}, vol.~10, p.~11, July 2023.

\bibitem{nyberg_echocardiographic_2023}
J.~Nyberg, E.~O. Jakobsen, A.~Østvik, E.~Holte, S.~Stølen, L.~Lovstakken, B.~Grenne, and H.~Dalen, ``Echocardiographic {Reference} {Ranges} of {Global} {Longitudinal} {Strain} for {All} {Cardiac} {Chambers} {Using} {Guideline}-{Directed} {Dedicated} {Views},'' {\em JACC: Cardiovascular Imaging}, p.~S1936878X23004163, Nov. 2023.

\bibitem{dosovitskiy_image_2021}
A.~Dosovitskiy, L.~Beyer, A.~Kolesnikov, D.~Weissenborn, X.~Zhai, T.~Unterthiner, M.~Dehghani, M.~Minderer, G.~Heigold, S.~Gelly, J.~Uszkoreit, and N.~Houlsby, ``An {Image} is {Worth} 16x16 {Words}: {Transformers} for {Image} {Recognition} at {Scale},'' June 2021.
\newblock arXiv:2010.11929.

\bibitem{asvold_cohort_2023}
B.~O. Åsvold, A.~Langhammer, T.~A. Rehn, G.~Kjelvik, T.~V. Grøntvedt, E.~P. Sørgjerd, J.~S. Fenstad, J.~Heggland, O.~Holmen, M.~C. Stuifbergen, S.~A.~A. Vikjord, B.~M. Brumpton, H.~K. Skjellegrind, P.~Thingstad, E.~R. Sund, G.~Selbæk, P.~J. Mork, V.~Rangul, K.~Hveem, M.~Næss, and S.~Krokstad, ``Cohort {Profile} {Update}: {The} {HUNT} {Study}, {Norway},'' {\em International Journal of Epidemiology}, vol.~52, pp.~e80--e91, Feb. 2023.

\bibitem{morris_normal_2017}
D.~A. Morris, M.~Krisper, S.~Nakatani, C.~Köhncke, Y.~Otsuji, E.~Belyavskiy, A.~K. Radha~Krishnan, M.~Kropf, E.~Osmanoglou, L.-H. Boldt, F.~Blaschke, F.~Edelmann, W.~Haverkamp, C.~Tschöpe, E.~Pieske-Kraigher, B.~Pieske, and M.~Takeuchi, ``Normal range and usefulness of right ventricular systolic strain to detect subtle right ventricular systolic abnormalities in patients with heart failure: a multicentre study,'' {\em European Heart Journal. Cardiovascular Imaging}, vol.~18, pp.~212--223, Feb. 2017.

\bibitem{wang_defining_2021}
T.~K.~M. Wang, R.~A. Grimm, L.~L. Rodriguez, P.~Collier, B.~P. Griffin, and Z.~B. Popović, ``Defining the reference range for right ventricular systolic strain by echocardiography in healthy subjects: {A} meta-analysis,'' {\em PLoS ONE}, vol.~16, p.~e0256547, Aug. 2021.

\bibitem{salte_artificial_2021}
I.~M. Salte, A.~Østvik, E.~Smistad, D.~Melichova, T.~M. Nguyen, S.~Karlsen, H.~Brunvand, K.~H. Haugaa, T.~Edvardsen, L.~Lovstakken, and B.~Grenne, ``Artificial {Intelligence} for {Automatic} {Measurement} of {Left} {Ventricular} {Strain} in {Echocardiography},'' {\em JACC: Cardiovascular Imaging}, vol.~14, pp.~1918--1928, Oct. 2021.

\bibitem{salte_deep_2023}
I.~M. Salte, A.~Østvik, S.~H. Olaisen, S.~Karlsen, T.~Dahlslett, E.~Smistad, T.~K. Eriksen-Volnes, H.~Brunvand, K.~H. Haugaa, T.~Edvardsen, H.~Dalen, L.~Lovstakken, and B.~Grenne, ``Deep {Learning} for {Improved} {Precision} and {Reproducibility} of {Left} {Ventricular} {Strain} in {Echocardiography}: {A} {Test}-{Retest} {Study},'' {\em Journal of the American Society of Echocardiography}, vol.~36, pp.~788--799, July 2023.

\end{thebibliography}

\end{document}